\documentclass[a4paper,11pt,twoside]{article}

\author{\large \bf G\"oran Einarsson \\ \\
         Royal Institute of Technology, Telecommunication Theory, \\
         Dept. of Signals, Sensors and Systems, 10044 Stockholm, Sweden. \\
         Phone;  +46 8 7906578, Fax: +46 8 7909370, 
         E-mail: einarsson@s3.kth.se}

\title{\Large \sc Quantum Theory and Classical Information}
\date{February 15, 2002}

\usepackage{psfrag}
\usepackage{amssymb}
\usepackage{amsmath}
\usepackage{graphicx}
\usepackage{longtable}

\begin{document}
\maketitle
\rule{0mm}{5mm} \\
{\sf Where there is quantum theory there is hope}\\
{\small Quoted from Joyce Carol Oates {\em What I lived for}}

\section*{Abstract}
Transmission of classical information 
using quantum objects such as polarized photons is
studied. The classical (Shannon) channel capacity and its relation to
quantum (von Neumann) channel capacity is investigated for various
receiver arrangements.

A quantum channel with transmission impairment caused by attenuation and 
random polarization noise is considered.
It is shown that the maximal (von Neumann) capacity of such a channel can
be realized by a simple symbol by symbol detector followed by a classical 
error correcting decoder.

For an intensity limited optical channel capacity is achieved by
on-off keying (OOK). The capacity per unit cost
is shown to be 1 nat/photon = 1.44 bit/photon, slightly larger than the
1 bit/photon obtained by orthogonal quantum signals.

\section{Introduction}
In his fundamental work \cite{Shannon 1948} {\em A Mathematical Theory 
of Communications}
from 1948 Claude Shannon introduced the quantity
\begin{displaymath}
                  H = - \sum p_i \log p_i
\end{displaymath}
which he called ``entropy''. It plays a central role
in information theory as a measure of information, choice 
and uncertainty. 

Richard P. Feynman tells \cite{Feynman} that Shannon adopted this
term on the advice of
John von Neumann, who declared that it would give him `` ... a great
edge in debates because nobody really knows what entropy is anyway''.

At MIT in the early 1960-ties Claude Shannon told me and the
other students in Course 6.575 that he choose the name because his
expression had the same form as that of entropy in statistical mechanics.
He also said that he doubted that information theory has
any physical relation to thermodynamics. 
\begin{figure}[t!]
   \begin{center}
   \includegraphics[width=1.0\textwidth]{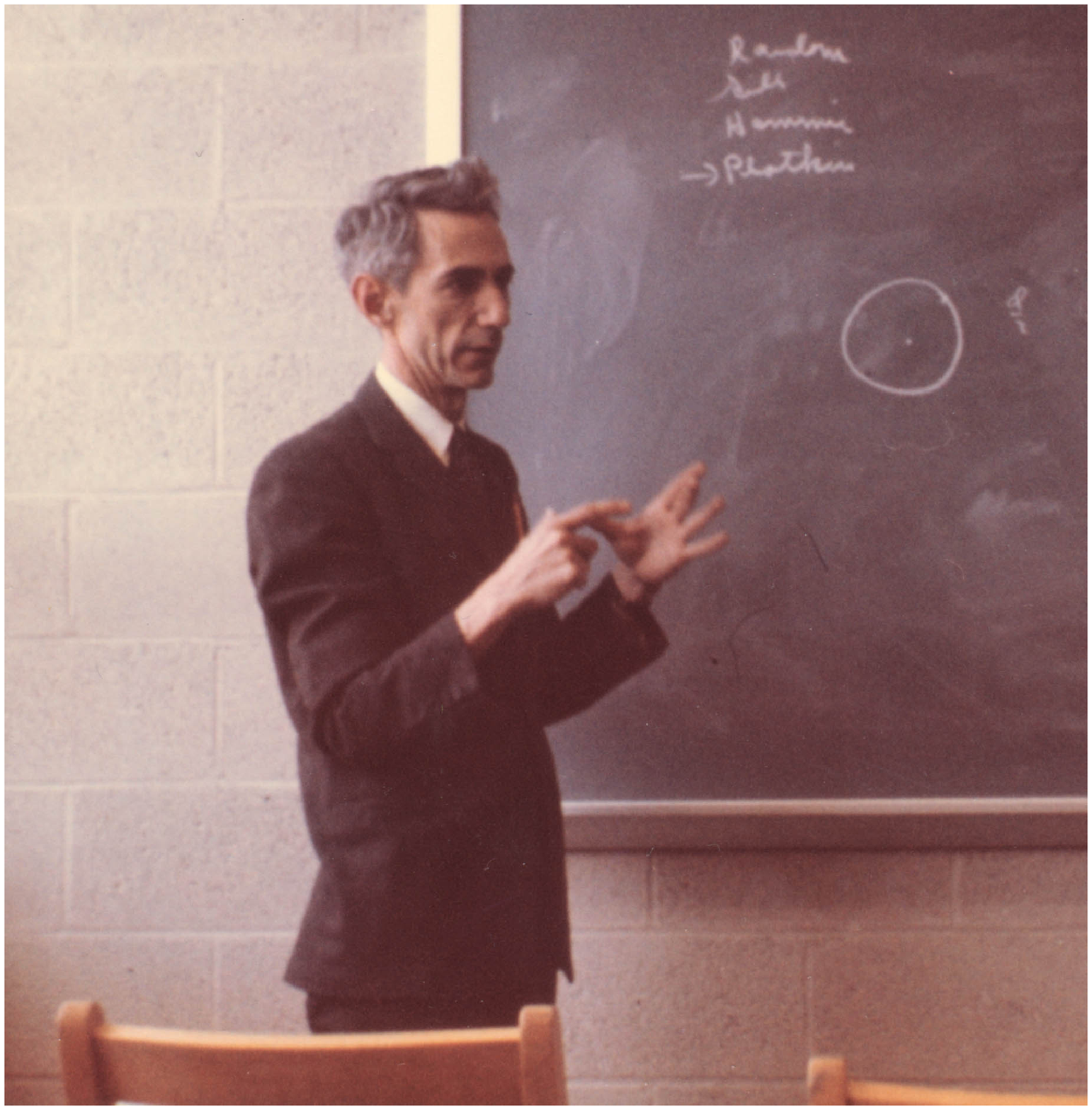}
   \end{center}
        \caption{ }
        \label{fig 0.1}
         Claude E. Shannon, MIT Course 6.575, April 17, 1961. \\
         {\small \sl Photo: G\"{o}ran Einarsson} \protect\\ 
\end{figure}
\newpage
One of the principal works on quantum theory is
{\em Mathematische Grundlagen der Quantenmechanik} \cite{von Neumann 1932}
by Johann von Neumann published in Berlin 1932, before he changed 
his first name to John.   
There he introduced the concept of quantum entropy
\begin{displaymath}
                     S = - {\rm Spur}(U \ln U)
\end{displaymath}
The motivation was of course thermodynamics, there was no information
theory around at that time.

More recently it has been shown by A. S. Holevo and others, 
that the von Neumann entropy S plays
a fundamental role in quantum information theory just as
the Shannon entropy H in classical information theory.

Quantum Information Theory is an interesting and expanding field.
The basic applications of classical information theory such as
source coding, data compression and channel coding have
counter parts in quantum theory. Most of today research is focused
on coding and transmission of quantum states motivated by
the connection to quantum computing.

We deal here with 
transmission of classical information by quantum objects.

\section{Basic Quantum Theory}

A pure quantum state is represented by a normalized vector in a
Hilbert space. We consider polarized photons in
a two-dimensional space. \\
A `ket' $|\psi\rangle$ is a column vector
\begin{displaymath}
        |\psi\rangle = \left[ \begin{array}{c}
                                      a   \\
                                      b    \\
                               \end{array} \right] 
\end{displaymath}
with complex components $a$ and $b$ and normalization
\begin{displaymath}
              \Vert\psi\Vert = |a|^2 + |b|^2 = 1
\end{displaymath}
The `bra'  $\langle\psi|$ is the complex transpose of $|\psi\rangle$
i. e. the row vector
\begin{displaymath}
        \langle\psi| = \left[ \begin{array}{cc}
                                      a^* & b^*  
                               \end{array} \right] 
\end{displaymath}
An important feature of a Hilbert space is the scalar product
$\langle\psi|\varphi\rangle$ called the `braket'. For
\begin{displaymath}
        |\psi\rangle = \left[ \begin{array}{c}
                                      a   \\
                                      b    \\
                               \end{array} \right]  \ \mathrm{and} \ \
        |\varphi\rangle = \left[ \begin{array}{c}
                                      c   \\
                                      d    \\
                               \end{array} \right]  
\end{displaymath}
the scalar product is
\begin{displaymath}
      \langle\psi|\varphi\rangle = \langle\varphi|\psi\rangle^* =
      a^*c + b^*d = (c^*a + d^*b)^*
\end{displaymath}
A two-dimensional quantum state $|\psi\rangle$ representing one bit
of information  and is called a {\em qubit}.

Consider polarized photons an let $|0\rangle$
denote the state of horizontal and $|1\rangle$
vertical polarization. These two states are orthogonal, i. e.
their scalar product is equal to zero, and
arbitrary polarization states can be expressed as a weighted
sum of these. The state 
\begin{displaymath}
      |\psi_1\rangle = {\scriptstyle \frac{1}{\sqrt2}}
           (|0\rangle +|1\rangle) 
\end{displaymath}
denotes 45 degree polarization and 
\begin{displaymath}
      |\psi_2\rangle = {\scriptstyle \frac{1}{\sqrt2}}
           (|0\rangle +j |1\rangle) 
\end{displaymath}
is right hand side circular polarization.

The scalar product between states plays an important role in
the sequel. By direct calculations it is easily  shown that
$\langle\psi_1|0\rangle = 1/\sqrt2$ and
$\langle\psi_2|1\rangle = j/\sqrt2$.
 
 \section{Communication of Classical Information}
\subsection{Binary Signaling}

We consider the possibility of communicating classical information,
i.e. ordinary data expressed as binary digits `one' and `zero'
utilizing quantum objects such as polarized photons.

\begin{figure}[tb]
   \begin{center}
   \leavevmode 
   \begin{psfrags}
     \psfrag{x}{\bf \Large $|0\rangle$}
     \psfrag{y}{\bf \Large $|1\rangle$}
     \psfrag{psi0}{\Large \bf $\chi_0$}
     \psfrag{psi1}{\bf \Large $\chi_1$}
     \psfrag{s0}{\bf \Large $|s_0\rangle$}
     \psfrag{s1}{\bf \Large $|s_1\rangle$}
     \psfrag{ss0}{\Large $s_0$}
     \psfrag{ss1}{\Large $s_1$}
     \psfrag{0}{\bf \Large $0$}
     \psfrag{1}{\bf \Large $1$}
     \psfrag{a}{\Large \bf a)}
     \psfrag{b}{\Large \bf b)}
     \psfrag{1.}{\large $1$}
     \psfrag{1/2}{ $1/2$}
     \psfrag{Q0}{$\scriptstyle Q(0)=3/5$}
     \psfrag{Q1}{$\scriptstyle Q(1)=2/5$}
     \psfrag{C_S}{$\pmb{C_S} = 0.32$}
     \psfrag{C_N}{$\pmb{C_N} = 0.60$}
     \includegraphics[width=1.0\textwidth]{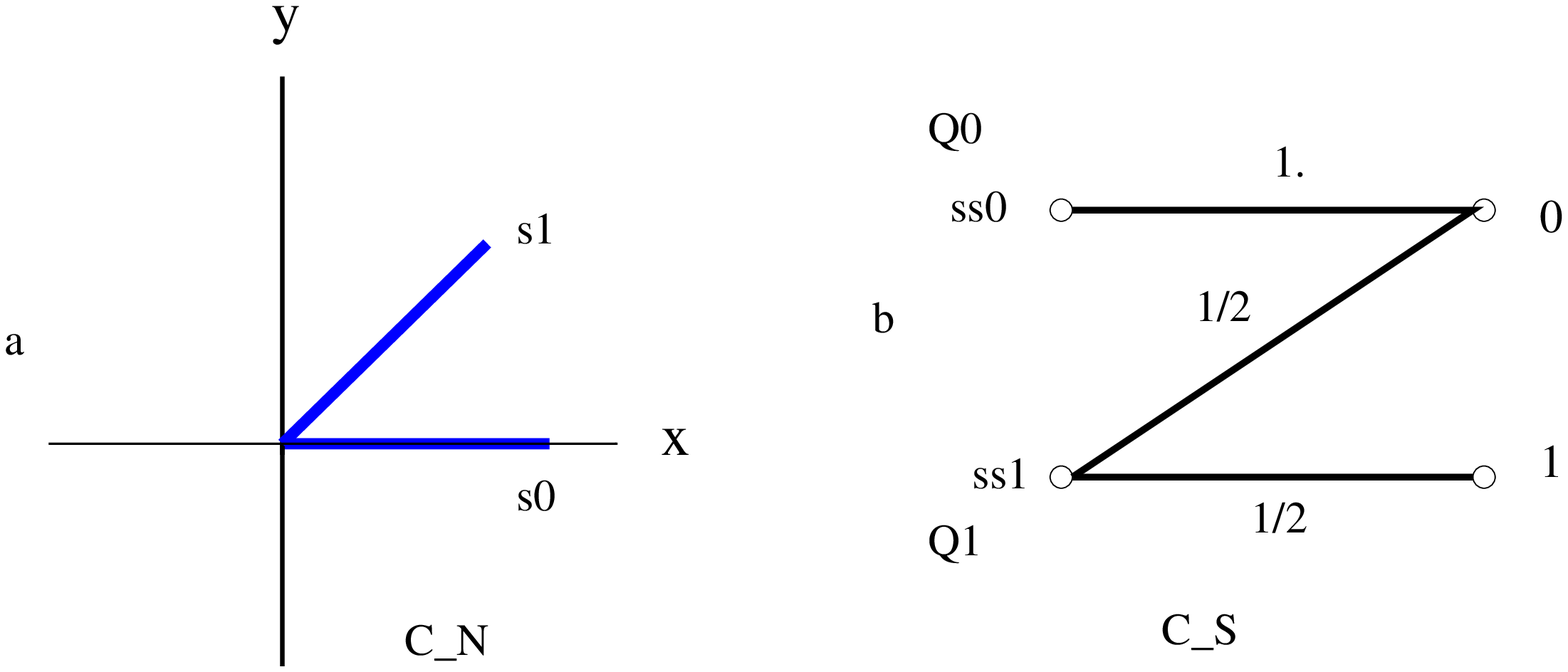}
   \end{psfrags}
   \end{center}
        \caption{ }
        \label{fig 1}
         Binary photon communication with horizontal
         polarization receiver.  \protect\\ 
         a) Transmitted states $|s_0 \rangle$ and $|s_1 \rangle$. \protect\\
         b) Resulting discrete binary Z-channel.
\end{figure}

As an example let the transmitter produce photons in two   
polarization states $|s_0\rangle$ and $|s_1\rangle$ shown in Fig. \ref{fig 1}.
\begin{displaymath}
              |s_0\rangle = |0\rangle, \ \ \ 
              |s_1\rangle = {\scriptstyle \frac{1}{\sqrt2}}
                            (|0\rangle + |1\rangle) 
\end{displaymath}
The transmitted signals have either horizontal 
or 45 degree polarization.

The receiver determines which type of photon was sent
by a suitable measuring device (receiver).
As a first attempt let the receiver consist 
of a horizontally oriented polarization filter.

A photon in state $|s_0\rangle$ will pass the receiver filter with 
certainty while a photon in state $|s_1\rangle$ will pass with probability
\begin{displaymath}
             |\langle s_1|0\rangle|^2 = 0.5
\end{displaymath}
This means that the communication system is equivalent to the discrete 
binary Z-channel, shown in Fig. \ref{fig 1}.

The Shannon channel capacity of a memoryless discrete channel 
with input alphabet $k=0,1,\dots, K-1$ and output alphabet 
$j=0,1,\dots,J-1$ is equal to the 
maximum of the average mutual information between sender and receiver
\cite{Gallager 1968}
\begin{equation}
\label{eq 3.1.1}
                     C_S  = \left. \begin{array}{c} \\
                                     \mathrm{max}    \\
                                     \scriptstyle{\{Q(k)}\} \\ 
                               \end{array} \right. 
                       \sum_{k,j} Q(k) P(j/k)
                       \log \frac{P(j/k)}{\sum_i Q(i)P(j/i)}
\end{equation}
The quantity $Q(k)$ denotes the  probability of input symbol $k$
and $P(j/k)$ is the transition probability between input symbol $k$
and output symbol $j$. Channel capacity defined as $I(X;Y)$ maximized
over all possible input symbol probability assignments
For a Z-channel with two input and output symbols $C$
can be obtained by a straightforward optimization. A convenient way
of evaluating  $C$ is to use the general 
expression for two-dimensional
channels presented in \cite{Nakagawa 1998}. 
For the Z-channel in 
Fig. \ref{fig 1} the optimal input distribution is $Q(0)=3/5$ and $Q(1)=2/5$
reflecting that the input symbol $k=0$ is more reliable than the
symbol $k=1$ and should be used more frequently in the code 
achieving capacity. The numerical value
is 
\begin{displaymath}
                       C_S =0.32\ \mathrm{bit/photon} 
\end{displaymath}

\begin{figure}[tb]
   \begin{center}
   \leavevmode 
   \begin{psfrags}
     \psfrag{x}{\bf \large $|0\rangle$}
     \psfrag{y}{\bf \large $|1\rangle$}
     \psfrag{psi0}{\large \bf $\chi_0$}
     \psfrag{psi1}{\bf \large $\chi_1$}
     \psfrag{s0}{\bf \large $|s_0\rangle$}
     \psfrag{s1}{\bf \large $|s_1\rangle$}
     \psfrag{ss0}{\Large $s_0$}
     \psfrag{ss1}{\Large $s_1$}
     \psfrag{0}{\bf \large $0$}
     \psfrag{1}{\bf \large $1$}
     \psfrag{a}{\large \bf a)}
     \psfrag{b}{\large \bf b)}
     \psfrag{.85}{\large $0.85$}
     \psfrag{.15}{ $0.15$}
     \psfrag{Q0}{$\scriptstyle Q(0)=1/2$}
     \psfrag{Q1}{$\scriptstyle Q(1)=1/2$}
     \psfrag{C_S}{$\pmb{C_S} = 0.40$}
     \psfrag{C_N}{$\pmb{C_N} = 0.60$}
     \includegraphics[width=1.0\textwidth]{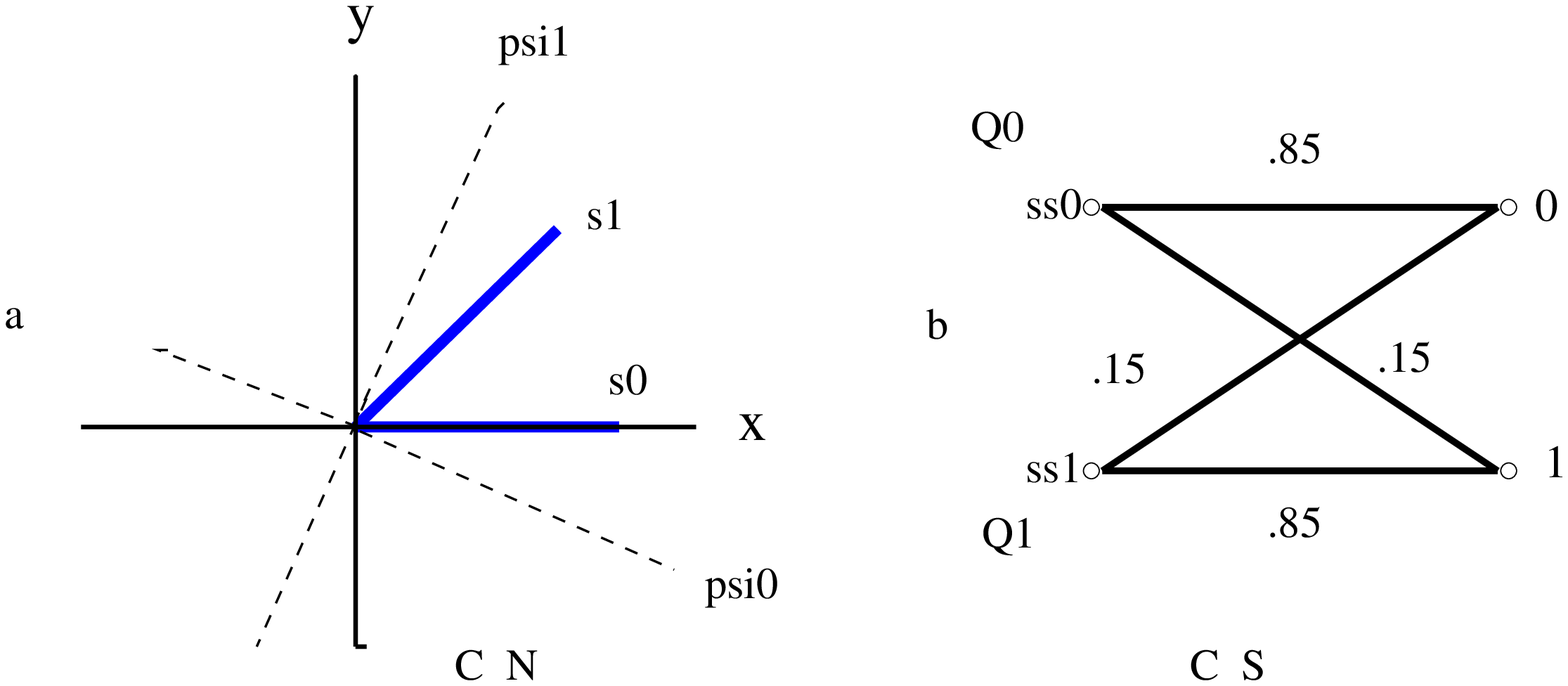}
   \end{psfrags}
   \end{center}
        \caption{ }
        \label{fig 2}
         Binary photon communication with optimal
         polarization receiver.  \protect\\
                 a) Transmitted states $|s_0 \rangle$ and 
                    $|s_1 \rangle$. Receiver orientation 
                    $\chi_0$ and $\chi_1$. \protect\\
                 b) Resulting discrete binary symmetric channel.
\end{figure}

A relevant question is whether the horizontally oriented polarization 
receiver is the best possible choice. There is a better
alternative. It has been shown \cite{Helstrom 1976} that a the filter
orientation shown in Fig. \ref{fig 2} minimizes the  probability of
making an incorrect decision.
In this case it also maximizes the Shannon capacity. 
The resulting discrete channel is symmetric with
\begin{displaymath}
                C_S = 1 - H(p) = 0.40\ \mathrm{bit/photon} 
\end{displaymath}
where $p$ is the transition probability and
\begin{equation}
\label{eqn 3.1.1.1}
              H(p)=- p \log p - (1-p) \log (1-p)
\end{equation}
is the binary entropy function.

\subsubsection{POVM Receiver}
\label{POVM rec}

The discrete classical channels arrived at in the preceding
section depend on the receiver configuration.
The receivers in Fig. \ref{fig 1} -- \ref{fig 2} perform simple
quantum tests, they check if the received photon is in any of two orthogonal
polarizations, which is the best that can be done operating in
an isolated two dimensional Hilbert space.

A more general type of measurement is a  POVM (Positive Operator
Valued Measure). In the present context it is accomplished 
by extending the original two-dimensional Hilbert space 
combining the received photon with a so called ancilla which is a photon in
a known fixed state. The received photon and the ancilla represent
four dimensions together, which makes a test between  four orthogonal states 
possible. This way it is possible to 
test if $s_0$ or $s_1$ were transmitted.
Such a decision can not be made with certainty and the
receiver will now also produce a no decision output.

\begin{figure}[tb]
   \begin{center}
   \leavevmode 
   \begin{psfrags}
     \psfrag{x}{\bf \large $|0\rangle$}
     \psfrag{y}{\bf \large $|1\rangle$}
     \psfrag{s0}{\bf \large $|s_0\rangle$}
     \psfrag{s1}{\bf \large $|s_1\rangle$}
     \psfrag{s0o}{\bf \large $|{\overline s}_0\rangle$}
     \psfrag{s1o}{\bf \large $|{\overline s}_1\rangle$}
     \psfrag{a}{\bf \large $|a\rangle$}
     \psfrag{b}{\bf \large $|b\rangle$}
     \psfrag{c}{\bf \large $|c\rangle$}
     \psfrag{e}{\large $\varepsilon$}
     \psfrag{1-e}{\large $1-\varepsilon$}
     \psfrag{A}{\Large $a$}
     \psfrag{B}{\Large $b$}
     \psfrag{C}{\Large $c$}
     \psfrag{aa}{\large \bf a)}
     \psfrag{bb}{\large \bf b)}
     \psfrag{S0}{\Large $s_0$}
     \psfrag{S1}{\Large $s_1$}
     \psfrag{CC}{$\pmb{C} = 0.29$}
     \includegraphics[width=1.0\textwidth]{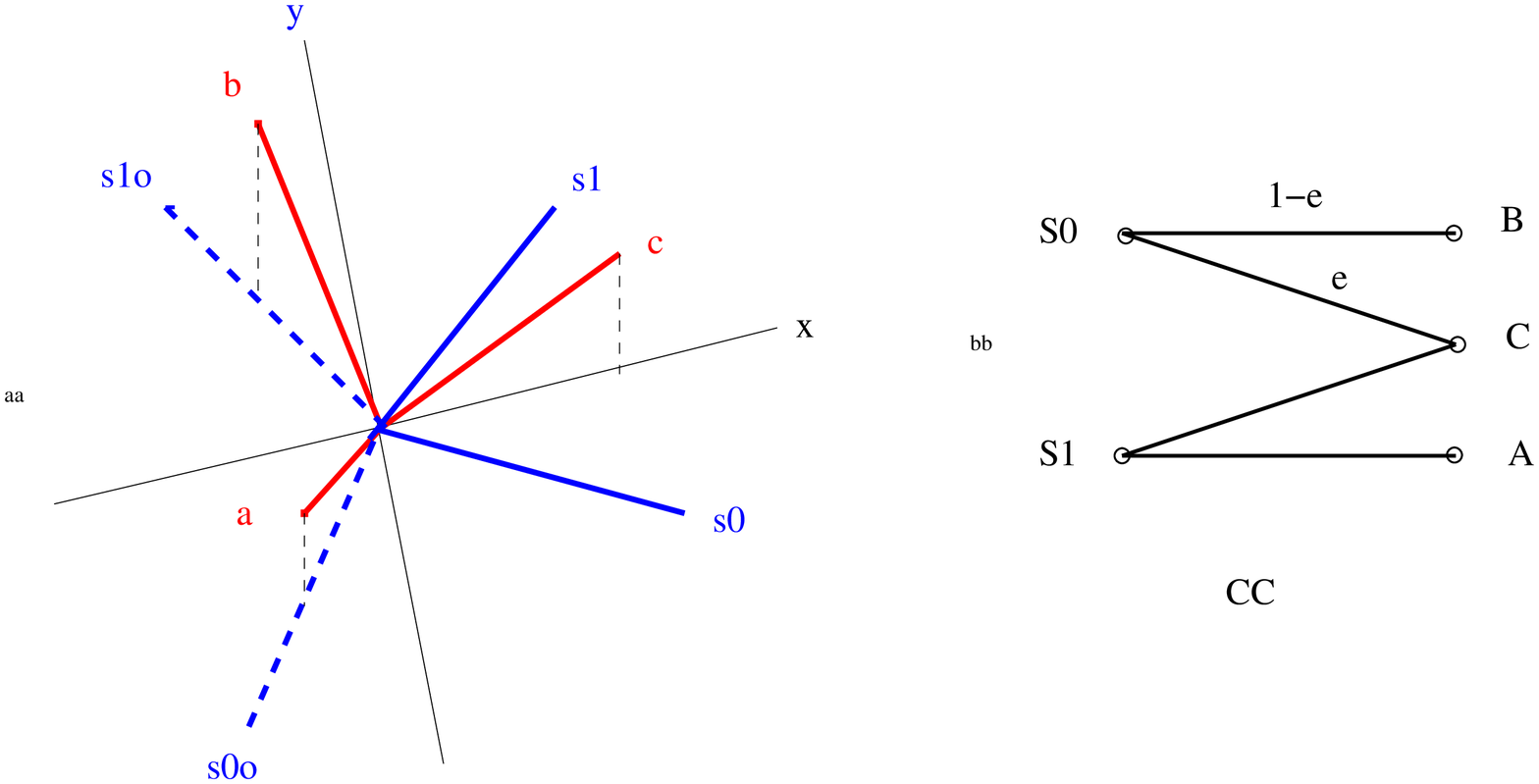}
   \end{psfrags}
   \end{center}
        \caption{ }

        \label{fig 3}
        \begin{tabbing}
        \=xx\=                                 \kill
        \>Binary erasure channel generated by 
        a POVM receiver. \protect\\
                 \>a) \>Transmitted states $|s_0 \rangle$ and $|s_1 \rangle$ 
                        together with the orthogonal \protect\\
             \> \>  POVM states $|a \rangle$, $|b \rangle$ and $|c \rangle$ 
                    in three dimensional space. \protect\\
                 \>b) \>Resulting discrete binary erasure channel,
                        $\varepsilon = 1/\sqrt{2}$.
        \end{tabbing}
\end{figure}

The principle of a POVM receiver for
two signals separated by 45 degrees
is illustrated in Fig. \ref{fig 3}.
The idea is to create three orthogonal state vectors
$|a\rangle$, $|b\rangle$ and $|c\rangle$ in the extended Hilbert space,
such that
the projections of two of them falls on 
the signal vectors $|s_0\rangle$ and $|s_1\rangle$. In the present
situation it is not possible to project directly on these and
the projections are made on the vectors  $|\bar{s}_0\rangle$ and 
$|\bar{s}_1\rangle$ orthogonal to  $|s_0\rangle$ and $|s_1\rangle$.
From the geometry follows 
\begin{displaymath}
              |\langle a|\bar{s}_0\rangle|^2 
              = |\langle b|\bar{s}_1\rangle|^2
              = \frac{1}{1 + \cos \beta} 
\end{displaymath}
with $\beta = \pi/4.$
The output symbol probabilities when, say $s_0$, is transmitted are
given by the square correlations between $|s_0\rangle$ and
$|a\rangle$, $|b\rangle$ and $|c\rangle$ respectively.
\begin{displaymath}
\left.
\begin{array}{lcl}
       P(a|s_0) & = & |\langle a|s_0\rangle|^2 
                  = |\langle a|\bar{s}_0\rangle
                     \langle \bar{s}_0|s_0\rangle|^2 = 0 \\ 
       P(b|s_0) & = & |\langle b|s_0\rangle|^2 
                  = |\langle b|\bar{s}_1\rangle
                     \langle \bar{s}_1|s_0\rangle|^2 
                  = 1 - \cos \beta \\
       P(c|s_0) & = & 1 - P(b|s_0) = \cos \beta
\end{array} \right\}
\end{displaymath}
The receiver bases its decision on the projections
in the original two-dimensional space
making its measurements on the received polarized 

\subsection{von Neumann Capacity}

The receivers in Fig. \ref{fig 1} -- \ref{fig 3}  make decisions
on each received symbol separately, they perform hard decisions.
In general the channel capacity can be improved by making decisions
based on a sequence of received symbols.
Holevo showed \cite{Holevo 1973} that the capacity
for transmission of classical information over a quantum channel
is upper bounded by
\begin{equation}
\label{eqn 3.2.1}
                            C_N  = \left. \begin{array}{c} \\
                                     \mathrm{max}    \\
                                     \scriptstyle{\{q_k}\} \\ 
                               \end{array} \right. 
                        \{S(\rho) - \sum_{k=1}^{N} q_k S(\rho_k)\}
\end{equation}
The transmitter sends on of $N$ possible states characterized by their
density matrices $\rho(k)$. The states may be pure or mixed. State $k$
has input probability $q_k$ and $\rho$ is the average density matrix
$\rho = \sum_k q_k \rho_k$. The function $S(\rho)$ is the von Neumann
entropy
\begin{equation}
\label{eq 3.2.2}
                    S(\rho) = -\mathrm{trace} \{\rho \log \rho\}             
\end{equation}
It has recently been proved that the Holevo upper bound (\ref{eqn 3.2.1})
actually defines capacity, i.e. it can be achieved. This result is known
as the Holevo-Schumacher-Westmoreland (HSW) Theorem, see \cite{Nielsen 2000}.

For pure input states $S(\rho_k) = 0$ and (\ref{eqn 3.2.1})
reduces to
\begin{equation}
\label{eqn 3.2}
                     C_N  = \left. \begin{array}{c} \\
                                     \mathrm{max}    \\
                                     \scriptstyle{\{q_k}\} \\ 
                               \end{array} \right. S(\rho) 
\end{equation}
For the signal configuration in Fig. \ref{fig 1}
the density matrices are
\begin{displaymath}
              \rho_0 =|s_0 \rangle\langle s_0| = \left[ \begin{array}{cc}
                                                              1 & 0   \\
                                                              0 & 0   \\
                                                 \end{array} \right]
\end{displaymath}
and
\begin{displaymath}
              \rho_1 =|s_1\rangle\langle s_1| = \frac{1}{2}
                                                \left[ \begin{array}{cc}
                                                       1 & 1   \\  
                                                       1 & 1   \\
                                                 \end{array} \right]
\end{displaymath}
Equal input probabilities $q_0 = q_1 = \frac{1}{2}$ gives
\begin{displaymath}
\label{eqn 3.4}
              \rho=\frac{1}{2}(\rho_o + \rho_1) 
                  =\frac{1}{4}\left[ \begin{array}{cc}
                                      3 & 1   \\  
                                      1 & 1   \\
                                 \end{array} \right]
\end{displaymath}
Substitution into (\ref{eqn 3.2}) gives the von Neumann capacity
for communication with two photons differing 45 degrees
in polarization
\begin{displaymath}
                       C_N = 0.60 \ \mathrm{bit/photon} 
\end{displaymath}
The maximal value $C_N$ can achieve is $C_N = 1$ which is
obtained by orthogonal signals, e.g. $s_0$ = $|0 \rangle$ 
and $s_1$ = $|1 \rangle$. In this case $C_S = C_N$ and
the the limit of 1 bit per photon is reached in a trivial way.

\subsection{Ternary signaling}

\begin{figure}[tbhp]
   \begin{center}
   \leavevmode 
   \begin{psfrags}
     \psfrag{x}{\bf \large $|0\rangle$}
     \psfrag{y}{\bf \large $|1\rangle$}
     \psfrag{s1}{\bf \large $|s_1\rangle$}
     \psfrag{s2}{\bf \large $|s_2\rangle$}
     \psfrag{s3}{\bf \large $|s_3\rangle$}
     \psfrag{1}{\bf \large $1$}
     \psfrag{2}{\bf \large $2$}
     \psfrag{3}{\bf \large $3$}
     \psfrag{ss1}{\bf \Large $s_1$}
     \psfrag{ss2}{\bf \Large $s_2$}
     \psfrag{ss3}{\bf \Large $s_3$}
     \psfrag{a}{\Large $|a\rangle$}
     \psfrag{b}{\Large $|b\rangle$}
     \psfrag{c}{\Large $|c\rangle$}
     \psfrag{at}{\Large $|a\rangle$}
     \psfrag{bt}{\Large $|b\rangle$}
     \psfrag{ct}{\Large $|c\rangle$}
     \psfrag{2/3}{\scriptsize $2/3$}
     \psfrag{1/6}{\scriptsize $1/6$}
     \psfrag{1/2}{\scriptsize $1/2$}
     \psfrag{aa}{\large \bf a)}
     \psfrag{bb}{\large \bf b)}
     \psfrag{cc}{\large \bf c)}
     \psfrag{C_S}{$\pmb{C_S} = 0.33$}
     \psfrag{C_2}{$\pmb{C_S} = 0.58$}
     \psfrag{C_N}{$\pmb{C_N} = 1.0$}
     \includegraphics[width=1.0\textwidth]{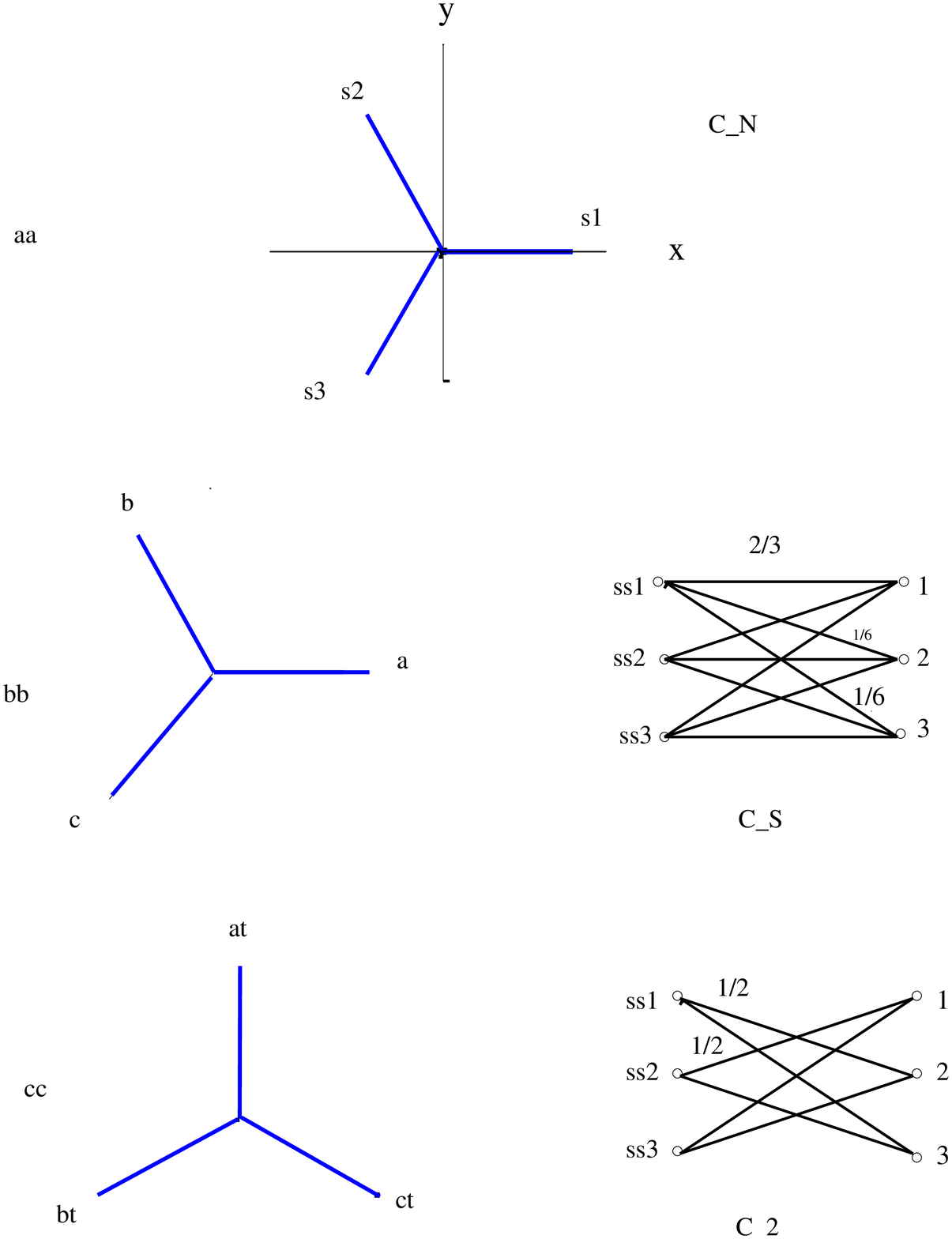}
   \end{psfrags}
   \end{center}
        \caption{ }
        \label{fig 4}
        Ternary signals.  \protect\\
                 a) Transmitted states $|s_1 \rangle$, 
                    $|s_2 \rangle$ and $|s_3 \rangle$. \protect\\
                 b) POVM vector projections parallel to the signals.
                    \protect\\
                 c) POVM vector projections orthogonal to the signals.
\end{figure}

There is no need to restrict the communication to binary transmission.
The following example of a ternary quantum signal alphabet
has been investigated by Peres and Wootters \cite{Peres 1991} and implemented
by Clarke et al. \cite{Clarke 2000}. The transmitted photon is on one of three
symmetrical polarizations 
$120^\circ$ apart shown in Fig. \ref{fig 4}a.
\begin{displaymath}
\left.
\begin{array}{lcl}
    |s_1 \rangle & = & \ \ |0 \rangle \\
    |s_2 \rangle & = & - {\scriptstyle \frac{1}{2}} |0\rangle +
                 {\scriptstyle \frac{\sqrt 3}{2}} |1\rangle)  \\
    |s_3 \rangle & = & - {\scriptstyle \frac{1}{2}} |0\rangle -
                 {\scriptstyle \frac{\sqrt 3}{2}} |1\rangle)  
\end{array} \right\}
\end{displaymath}
The von Neumann capacity (\ref{eqn 3.2}) for this signal set is   
equal to the maximally possible  $C_N =1$.

A POVM receiver utilizing an ancilla photon, analogous to the one 
described in Section \ref{POVM rec}
is needed to be able to distinguish between the three transmitted
alternatives. A natural choice is to let POVM projections
fall on the input signals, which would correspond to a matched filter
receiver in classical communication theory. The resulting
discrete channel is shown in Fig. \ref{fig 4}b. Its
Shannon capacity is $C_S =0.33$ bit/photon. This receiver
maximizes the probability of detecting the correct signal.

A better alternative, however, is to let the POVM projections
be orthogonal to the input signal. This results in the channel
in Fig. \ref{fig 4}c with  $C_S =0.58$ bit/photon.

The Shannon capacity requires a maximization
over the input symbol alphabet. One possibility is to
refrain from the use of one of the
input symbols, i.e. assign probability zero to it. In the present
case the result is a binary channel with 
two signals separated by $120^\circ$ or equivalently $60^\circ$
in polarization. An optimum polarization receiver of the type
illustrated in Fig. \ref{fig 2} generates
a binary symmetric channel with $p = 0.067$ and a capacity
$C_S = 0.65$ bit/photon. Which is a larger value than for the
ternary signaling systems above  and almost as good as the photon pair
receiver below.
\begin{figure}[b!]
   \begin{center}
   \leavevmode 
   \begin{psfrags}
     \psfrag{a}{\bf \large $|a\rangle$}
     \psfrag{b}{\bf \large $|b\rangle$}
     \psfrag{c}{\bf \large $|c\rangle$}
     \psfrag{s11}{\bf \large $|s_1s_1\rangle$}
     \psfrag{s22}{\bf \large $|s_2s_2\rangle$}
     \psfrag{s33}{\bf \large $|s_3s_3\rangle$}
     \psfrag{s1s1}{\bf \large $s_1s_1$}
     \psfrag{s2s2}{\bf \large $s_2s_2$}
     \psfrag{s3s3}{\bf \large $s_3s_3$}
     \psfrag{s2}{\bf \large $|s_2\rangle$}
     \psfrag{s3}{\bf \large $|s_3\rangle$}
     \psfrag{1}{\bf \large $1$}
     \psfrag{2}{\bf \large $2$}
     \psfrag{3}{\bf \large $3$}
     \psfrag{aa}{\large \bf a)}
     \psfrag{bb}{\large \bf b)}
     \psfrag{.97}{\scriptsize $0.97$}
     \psfrag{.15}{\scriptsize $0.015$}
     \psfrag{C_S}{$\pmb{C_S} = 1.37/2 = 0.68$}
     \psfrag{C_N}{$\pmb{C_N} = 1.5/2 = 0.75$}
     \includegraphics[width=1.0\textwidth]{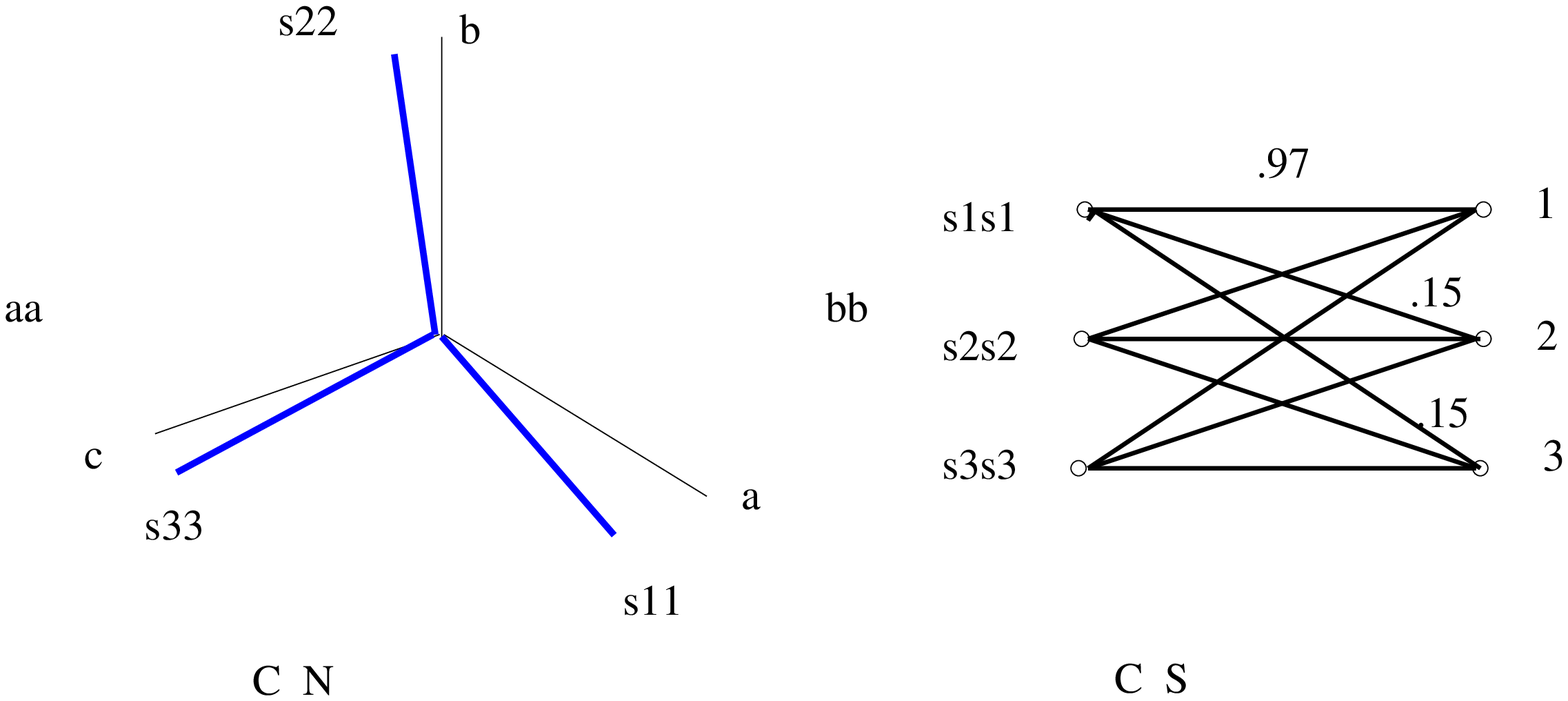}
   \end{psfrags}
   \end{center}
        \caption{ }
        \label{fig 5}
        \begin{tabbing}
        \=xx\=                                 \kill
        \>Ternary signals with photon pair transmission.  \protect\\
        \>a) \>Transmitted states and the orthogonal measurement  \protect\\
        \>   \>directions $|a \rangle$, $|b \rangle$ and $|c \rangle$ 
               in three dimensional space. \protect\\
        \>b) Resulting discrete ternary channel.
        \end{tabbing}
\end{figure}

An interesting alternative is to let the transmitted signals be
a pair of photons with equal polarization. 
The input alphabet is restricted to the three alternatives
$|s_1 s_1 \rangle$ $|s_2 s_2 \rangle$ and $|s_3 s_3 \rangle$. 
The von Neumann capacity for this signal set is  
$C_N$=1.5/2 = 0.75 bit/photon. 
It turns out that the transmitted state vectors are almost orthogonal in
four-dimensional Hilbert space, see Fig. \ref{fig 5}. 
The corresponding discrete ternary 
channel has $C_S$ = 1.37 corresponding to 0.68 bit/photon.
No ancilla photon arrangement is needed in this receiver.

\subsection{Quantum Cryptography}

Quantum communication is limited by the fundamental fact that
only orthogonal quantum states can be distinguished reliably.
In Quantum cryptography \cite{Bennett 1992} the fundamental 
uncertainty of the outcome
of a quantum measurement has been turned into an advantage.
Secret information can be communicated safe against eavesdropping. 

\subsection{Noisy channels}

So far we have assumed that an error free channels is available between
transmitter and receiver. In practice communication signals are subjected to
various types of impairment during transmission. In classical communication
noise of different origin is almost always present and has to be 
included in the analysis.

\subsubsection{Attenuation}
An common feature in communication systems is signal attenuation.
In quantum communication it means that photons or other signal
elements are lost on their
way from sender to receiver. Attenuation is usually expressed in dB
and if an average fraction $\varepsilon$ is lost it corresponds to
\begin{displaymath}
                    A = -10 \log(1-\varepsilon) \ \mathrm{dB}
\end{displaymath}
If the receiver works in synchronism with the transmitter it
can determine when photons are missing and such an event constitutes an
erasure. 
Erasures decrease the  capacity of a binary symmetric channel 
by a factor $1-\varepsilon$.
As an example let $\varepsilon$ = 10 \%, corresponding to
$A$ = 0.46 dB, the Shannon capacity for the $45^\circ$ binary
communication decreases from $C_S$ = 0.40 to 
$C_S$ = 0.36 bit/photon.
For a system using orthogonal polarizations, which would be error free
in case of perfect detection, an attenuation $A$ = 1 dB will reduce
the capacity from $C_S$ = 1  to $C_S$ = 0.79 bit/photon.

\subsubsection{Polarization noise}

A more complicated kind of impairment is if the polarization of the photon
is disturbed during transmission. If the polarization of the state
$|s_0 \rangle$ is changed by an angle $\varphi$
during transmission the received density matrix becomes
\begin{displaymath}
                  \rho_0(\varphi) = \left[ \begin{array}{cc}
                  \cos^2 \varphi & \cos \varphi \sin \varphi   \\
                  \sin \varphi \cos \varphi  & \sin^2 \varphi   \\
                                          \end{array} \right]
\end{displaymath}
When $\varphi$ is random the received state is a mixed state with
density matrix $\bar{\rho}_0 = \mathrm{E}\{\rho_0(\varphi)\}$.
For polarization noise with a  probability density $f(\varphi)$ 
symmetrical around $\varphi = 0$
\begin{equation}
\label{eqn 3.4.3}
                  \bar{\rho}_0 = \left[ \begin{array}{cc}
                          1 - d & 0   \\
                             0  & d    \\
                                     \end{array} \right]
\end{equation}
where
\begin{equation}
\label{eqn 3.4.4}
                 d = \int f(\varphi) \sin^2 \varphi \ d\varphi
\end{equation}
The constant $d$ depends on the shape of $f(\varphi)$. For a 
distribution  uniform between $-A/2$ and $A/2$ it is equal to
$d = (1 - (\sin A)/A)/2$ and for a Gaussian distribution
with variance $\sigma^2$ the quantity $d = (1 - \exp - 2 \sigma^2)/2$.

We consider communication with orthogonal signals $|s_0\rangle = |0\rangle$
and $|s_1\rangle = |1\rangle$. The received density matrix generated
by $\rho_1$ is
\begin{equation}
\label{eqn 3.4.5}
                  \bar{\rho}_1 = \left[ \begin{array}{cc}
                             d & 0   \\
                             0 & 1 - d    \\
                                     \end{array} \right]
\end{equation}
The average density matrix is
\begin{displaymath}
            \bar{\rho} = \frac{1}{2}(\bar{\rho}_0 + \bar{\rho}_1)
                       = \frac{1}{2} \left[ \begin{array}{cc}
                             1 & 0   \\
                             0 & 1   \\
                                     \end{array} \right]
\end{displaymath}
and the von Neumann entropy (\ref{eq 3.2.2}) is $S(\bar{\rho}) = S = 1$.
The entropies for (\ref{eqn 3.4.3}) and (\ref{eqn 3.4.5}) are
equal to $S_0 = S_1 = H(d)$ where $H(d)$ is the binary entropy
function (\ref{eqn 3.1.1.1}).
The von Neumann capacity (\ref{eqn 3.2.1}) is
\begin{displaymath}
\label{eqn 3.4.6}
               C_N = S - \frac{1}{2}(S_0 + S_1) = 1 - H(d)
\end{displaymath}
For a fixed polarization deviation the system generates 
a binary symmetric channel (BSC) with transition probability
$p(\varphi) = \sin^2 \varphi$. For a memoryless channel with
random $p$ the capacity is determined by the average value $\bar{p}$
equal to $d$ defined in (\ref{eqn 3.4.4}).
The Shannon capacity $C_S = 1 - H(\bar{p})$ is, for this noisy quantum 
channel, equal to its von Neumann capacity $C_N$. 

This means that a hard decision symbol by symbol receiver
in combination with a classical error correcting code 
is sufficient to achieve the ultimate capacity $C_N$ of this noisy
quantum channel.
The choice of orthogonal transmitted signals is essential for
this result.

As an example of the effect of polarization noise let $d = 0.1$.
The resulting channel capacities are $C_S$ = $C_N$ = 0.53 bit/photon.

%\newpage                         \hspace{100mm}   \today
\section{Alternative modulation}
\subsection{Pulse Position Modulation (PPM)}

Helstrom \cite{Helstrom 1974}  suggests a quantum modulation scheme using 
$M$ orthogonal states generated as longitudinal modes in
an ideal laser. The signaling is done by exciting one of these modes
into a coherent state.
It is shown that  the error probability goes to zero for increasing $M$,
which means that the channel have infinite capacity. He remarks
that `the quantum-mechanical nature of signals themselves does not
limit the information-carrying capacity of a coherent optical channel'.

To cast some light on this 
seemingly impossible  result consider
a semi-classical model of optical communication, the Poisson
channel. It models light as a random stream of photons characterized 
by the optical intensity $\gamma(t)$ \cite{Saleh 1978}. 
The number of photons in a
time interval $0-T$ has a Poisson distribution
\begin{displaymath}
        P(N=n) = \frac{m^n e^{-m}}{n!}
\end{displaymath}
with mean value 
\begin{displaymath}
            m = \int_0^T \gamma(t) \, \mathrm{d}t
\end{displaymath}
In  quantum terminology
the light is in a coherent state.

\begin{figure}[tbhp]
   \begin{center}
   \leavevmode 
   \begin{psfrags}
%     \psfrag{I}{\bf I}
     \psfrag{a}{\bf  a)}
     \psfrag{b}{\bf  b)}
     \psfrag{1}{\bf \large $1$}
     \psfrag{2}{\bf \large $2$}
     \psfrag{M}{\bf \large $M$}
     \psfrag{T}{$T$}
     \psfrag{t}{\bf \large $t$}
     \psfrag{g0}{\large \bf $\gamma_0$}
     \psfrag{g1}{\large \bf $\gamma_1$}
     \psfrag{1-e}{\large \bf $1-\varepsilon$}
     \psfrag{e}{\large \bf $\varepsilon$}
     \psfrag{er}{erasure}
     \includegraphics[width=1.0\textwidth]{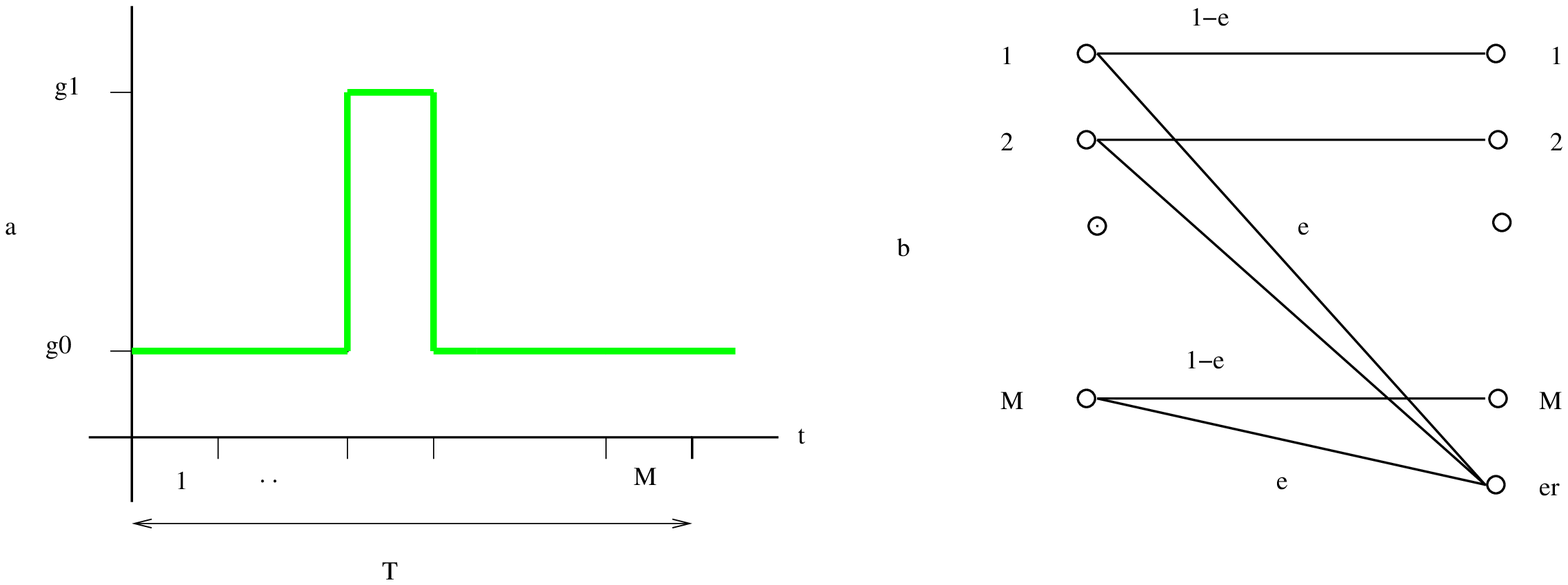}
   \end{psfrags}
   \end{center}
        \caption{ }
        \label{fig 6}
                 a) Optical pulse position modulation (PPM) with 
                    $M$ time slots.  \protect\\
                 b) Channel diagram for an ideal optical PPM channel
                    with $\gamma_0 = 0$.
\end{figure}

A set of $M$ orthogonal optical signals is generated
by dividing the transmission  interval of duration $T$  into $M$ time slots
each of width $ \Delta = M/T$. An pulse of limited optical energy $E$
is transmitted in one of the time slots. This modulation format 
called   Pulse Position Modulation (PPM)  is shown in Fig \ref{fig 6}a.
for the ideal case
with the background optical intensity $\gamma_0 =0$.
With this assumption, the only situation when a
transmission error can occur is when, due to the Poisson fluctuations, no
photons are observed in the interval $0-T$. 
The ideal PPM channel is equivalent to an M-ary erasure channel, shown in
Fig~\ref{fig 6}b.
The probability for an erasure is 
\begin{displaymath}
        P(N=0) =  e^{-m} = \varepsilon 
\end{displaymath}
where $m$ is the average number of transmitted photons.
The capacity for the M-ary erasure channel is
\begin{displaymath}
\label{eqn 4.2}
        C = (1 - \varepsilon) \log M \ {\rm bit/transmission}
\end{displaymath}
The channel capacity for optical PPM approaches $\infty$ 
when $M \to \infty$. It can be shown \cite{Davis 1980} that
this is true also when $\gamma_0 > 0$.
In practice the time slots can not be made arbitrary small since
optical bandwidth is limited. 
In PPM the optical pulses have optical intensity $E/\Delta$
which is assumed to be unlimited.

\subsection{On-Off Keying (OOK)}
For an optical system with constrained signal intensity $\gamma$ 
it has been proved \cite{Wyner 1988} that 
on-off keying (OOK) is the optimal modulation format. \\
With $\gamma \leq \gamma_1$ the
channel capacity, in natural units per second, is \cite{Einarsson 1989}, 
\cite{Wyner 1988}, \cite{Einarsson 1995}
\begin{equation}
\label{eqn 4.3}
          C = \frac{\gamma_0}{e} \left(\frac{\gamma_1}{\gamma_0}\right)^
                                    {\gamma_1/(\gamma_1 - \gamma_0)} 
            - \frac{\gamma_0 \gamma_1}{\gamma_1 - \gamma_0}
                           \ln\left(\frac{\gamma_1}{\gamma_0}\right) 
                                                    \ {\rm nat/s}
\end{equation}
The background optical intensity $\gamma_0 =0$ typically represents
the dark current in the receiver photo detector. \\
For the ideal case $\gamma_0 =0$ the expression reduces to
\begin{equation}
\label{eqn 4.4}
              C =\gamma_1/e  \ {\rm nat/s}
\end{equation}
The maximal error free transmission capability, channel capacity, is
usually expressed as information per unit time (bit/s or nat/s).  An alternative 
measure is capacity per unit cost \cite{Verdu 1990}.
For optical transmission a natural  cost function is the number of 
photons needed to reliably transmit one bit of information.

\begin{figure}[tb]
   \begin{center}quant-ph
   \leavevmode 
   \begin{psfrags}
     \psfrag{gamma1  photons/s}{$\gamma_1$ \ \ \ \ photons/s }
     \psfrag{cost/bit}{\large cost/bit}
     \psfrag{ln 2}{ln 2}
     \psfrag{gamma0=1}{$\gamma_0 = 1$}
     \includegraphics[width=1.0\textwidth]{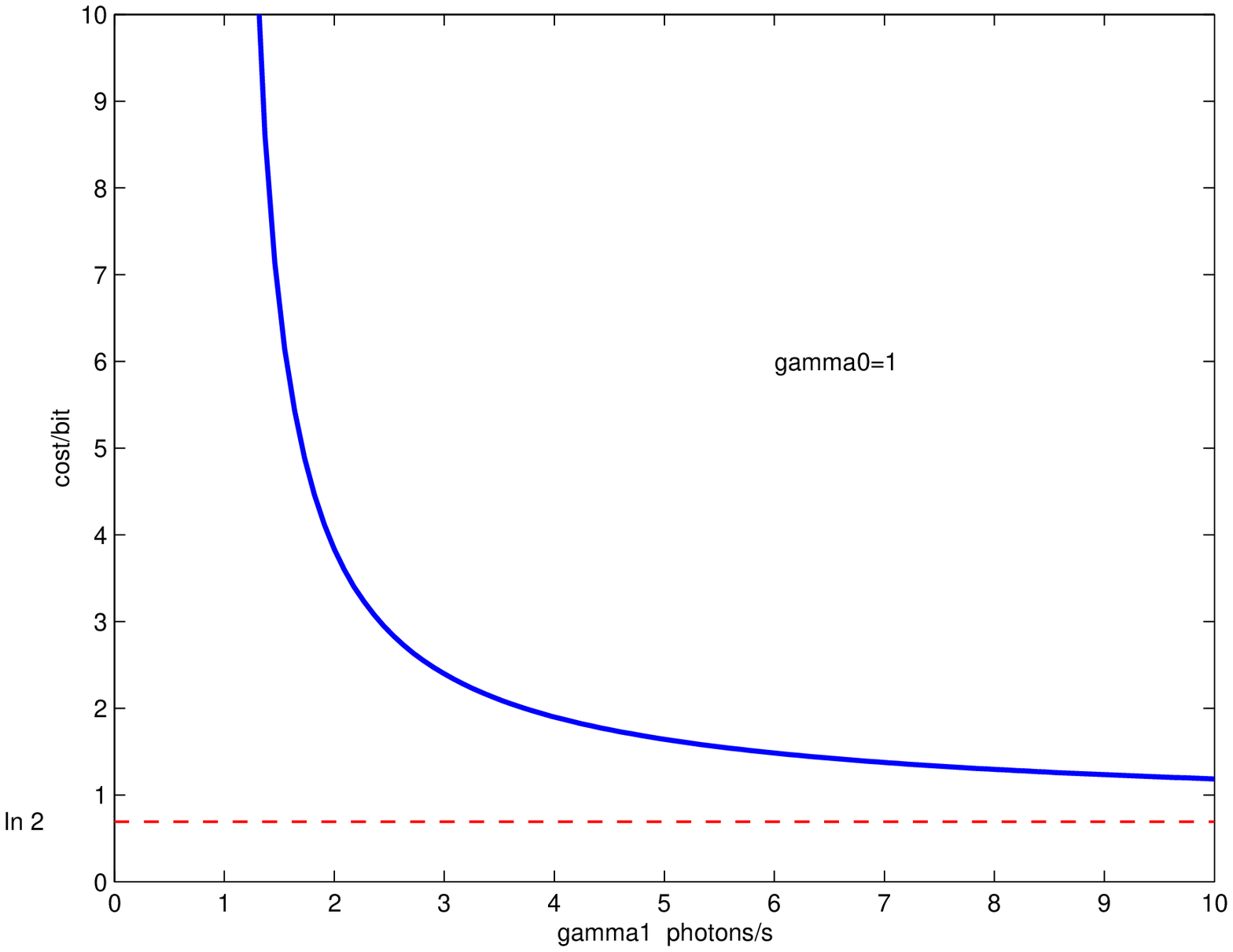}
   \end{psfrags}
   \end{center}
        \caption{ }
        \label{fig 7}
               Intensity limited binary optical system. \\
               Transmission efficiency expressed as unit cost
                (no. of photons) per bit.
\end{figure}

The capacity (\ref{eqn 4.3}) is achieved when the `on' symbol is
used with probability
\begin{equation}
\label{eqn 4.3.1}
        q = \frac{\gamma_0}{\gamma_1 - \gamma_0}
            \left[\frac{1}{e}\left(\frac{\gamma_1}{\gamma_0}\right)^
                                  {\gamma_1/(\gamma_1 - \gamma_0)} - 1 \right]
\end{equation}
The average intensity of signal photons is 
\begin{equation}
\label{eqn 4.3.2}
              \gamma_{ave} = q(\gamma_1-\gamma_0) \ \ {\rm photons/s}
\end{equation}
The capacity per unit cost in nat per photon becomes
\begin{equation}
\label{eqn 4.3.3}
              C_{ph} = C/\gamma_{ave} \ \ {\rm nat/photon}
\end{equation}
Fig \ref{fig 7} shows the cost per bit i.e. $\ln(2)/C_{ph}$ as a function of
$\gamma_1$ for a system with $\gamma_0 = 1$.
The diagram is equivalent to Fig 1 in \cite{Verdu 1990}
for an AWGN channel. For large $\gamma_1$ 
the capacity $C_{ph}$ approaches the asymptotic value 
             $1\ {\rm nat/photon} = 1.44\ {\rm bit/photon}$.
In Fig \ref{fig 7} the asymptotic value $\ln 2$ = 1/1.44 is indicated.
The symbol probability (\ref{eqn 4.3.1}) is $q = 1/{\rm e}$
for $\gamma_0 = 0$ and
\begin{equation}
\label{eqn 4.5}
              C_{ph} = C/q = 1\ {\rm nat/photon}
\end{equation}
for all values of $\gamma_1$.

The same limiting rate 1 nat/photon was obtained by J. R. Pierce 
\cite{Pierce 1978} for a receiver with an ideal optical amplifier.

The result (\ref{eqn 4.5}) can be obtained directly from a general
relation for channels with a zero-cost input symbol \cite{Verdu 1990}.
\begin{equation}
\label{eqn 4.5.1}
       C = \sup_X \frac{D(P_{Y \vert X=0} \Vert P_{Y \vert X=x})}{b[x]}
\end{equation}
where the supremum is over the input alphabet and $P_{Y \vert X=x}$
denotes the output distribtion given that the input is $x$. The quantity
$D(P \Vert Q)$ is the (Kullback-Liebler) divergence between the
probability distributions $P$ and $Q$, see \cite{Csiszar-Körner},
and $b[x]$ denotes the cost function,
\begin{equation}
\label{eqn 4.5.2}
         D(P \Vert Q) = \sum_{x \in X} P(x) \log \frac{P(x)}{Q(x)}             
\end{equation}
For the Z-channel is
\begin{equation}
\label{eqn 4.5.4}
     D = 1 \cdot \log \frac{1}{p} + 0 \cdot \log \frac{0}{1-p} 
       = -\log p = m  
\end{equation}
and
\begin{equation}
\label{eqn 4.5.5}
                 C = \frac{D}{b[x]} = \frac{m}{m} = 1\ {\rm nat/photon}
\end{equation}
Relation (\ref{eqn 4.5.1}) differs from Theorem 3 in
\cite{Verdu 1990} which seems to be incorrect.

\subsection{Limited Bandwidth}
The capacity (\ref{eqn 4.3}) is achieved when the pulse width $T$
goes to zero which means that an infinite
system bandwidth is required. 

%To illustrate that the the unlimited bandwidth assumption is 
%not essential to achieve efficiencies close to (\ref{eqn 4.5}) 

The following example illustrates that a transmission efficiency equal to 
(\ref{eqn 4.5}) can be achieved by a band limited system.

Consider on-off modulation with a finite pulse duration $T$
corresponding to a system bandwidth of the order of $R=1/T$.
For simplicity we consider an ideal system with $\gamma_0 = 0$
A receiver using symbol by symbol detection is equivalent to
a Z-channel with transition probability $p = e^{-m}$, 
c. f. Fig. \ref{fig 1}. 
The parmeter $m=\gamma_1 T$ is equal to the average number of 
photons in a received pulse.
\begin{figure}[tb]
   \begin{center}
   \leavevmode 
   \begin{psfrags}
   \psfrag{m   photons}{ $m$ \ \ \ photons}
   \psfrag{Cph  nat/photon}{$C_{ph}$ \ \ \ nat/photon}
     \includegraphics[width=1.0\textwidth]{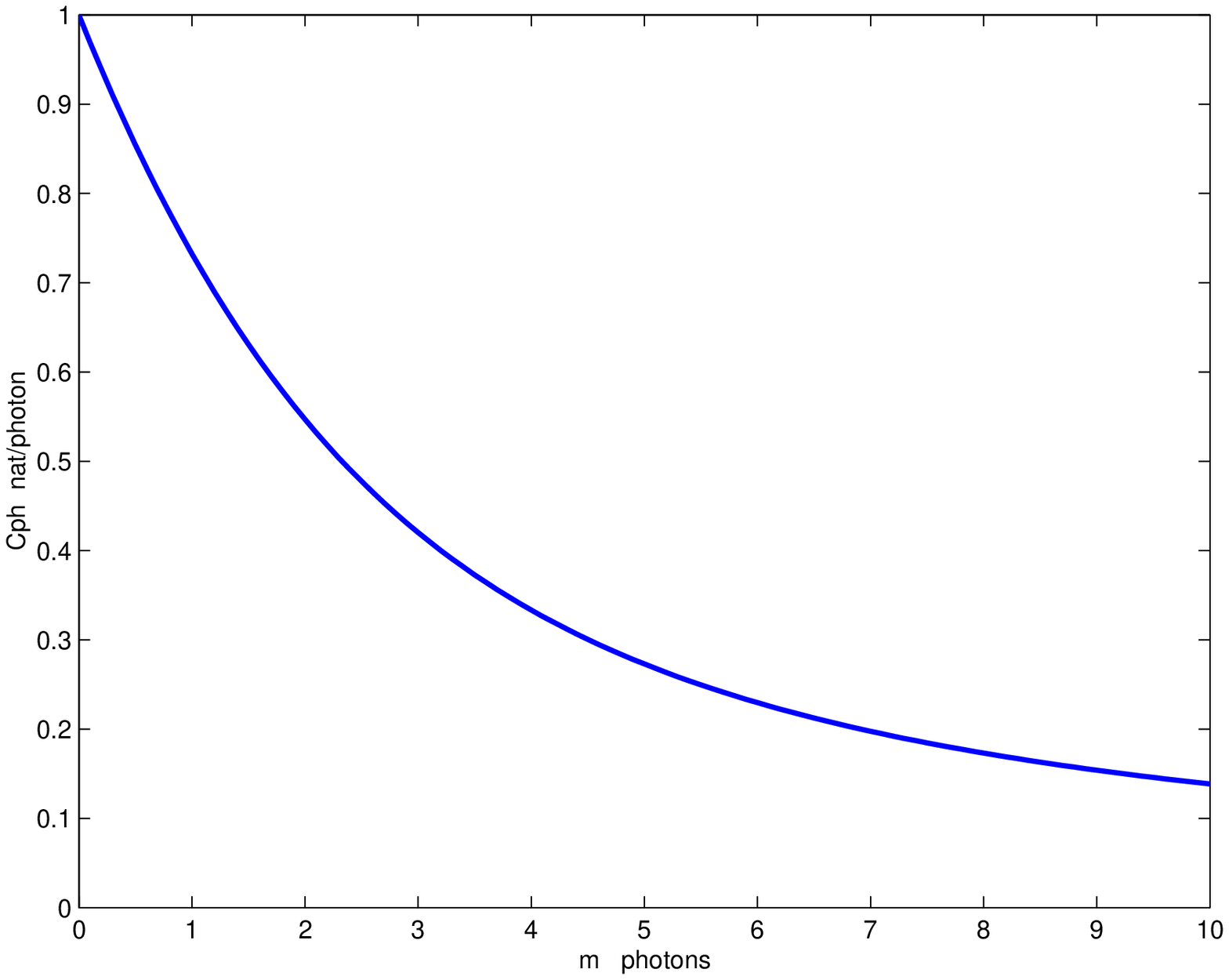}
   \end{psfrags}
   \end{center}
        \caption{ }
        \label{fig 8}
         Transmission capacity in nat per photon as a function of
         $m$ the average number of transmitted photons for an
         ideal ($\gamma_0 = 0$) band limited (OOK) optical channel.

\end{figure}
The mutual information between input and output for the channel is
\begin{equation}
\label{eqn 5.1}
     I(X;Y)=-q(1-p)\ln q + qp\ln p - (1-q(1-p))\ln(1-q(1-p)) 
\end{equation}
The channel capacity is the maximal value of (\ref{eqn 5.1})
which is achieved for 
\begin{equation}
\label{eqn 5.2}
  q = \frac{p^{(\frac{p}{1-p})}}{1+p^{(\frac{p}{1-p})}-p^{(\frac{1}{1-p})}}
\end{equation}
Fig \ref{fig 8} shows the capacity per (average) photon 
$C_{ph}=C/(q \cdot m)$ as a function of $m$. The efficiency is 
decreasing with $m$ and approaches its maximal value 
\begin{equation}
\label{eqn 5.2.1}
              C_{ph} = 1 \ {\rm nat/photon} 
\end{equation}
when $m$ goes to zero.

The result is independent of 
the symbol rate $R$ and thus of the system bandwidth. The capacity,
however, is low in terms of the symbol rate.

%The Z-channel capacity per photon (\ref{eqn 5.2.1}) can be derived 
%in a more direct way from Theorem 3 in \cite{Verdu 1990}.

%\newpage                         \hspace{100mm}   \today
\subsection{Entanglement-assisted communication}

A quantum property that has no counterpart in classical physics
is quantum correlation (entanglement). Two entangled photons
have features together that can not be attributed to the individual
photons. Two photon in an entangled state
\begin{displaymath}
      |\psi\rangle = {\scriptstyle \frac{1}{2}}
           (|00\rangle +|11\rangle) 
\end{displaymath}
will both be found in horizontal or both in vertical polarization 
when they are measured at separate locations.

Entanglement exists over arbitrary distances but it can not be used for
direct transmission of information. All information about an entangled pair
is contained in their joint density matrix, which is fixed from the
beginning. Whatever kind of operation  made on one of the
photons can not be detected by
any kind of measurement on the other photon.

Quantum correlation, however, can be used in combination with classical
communication in entanglement-assisted communication \cite{Bennett 1998}.
As an example communication of two bit of information between two parties 
Alice and Bob can be done in the following way. Alice prepare an
entangled pair and send one of the photons to Bob who stores it
unchanged. At a later time Alice operates on here photon
positioning the pair into one of four orthogonal entangled states.
She then sends her photon to Bob, with both photons available,
can determine which state was prepared 
and thereby decode two bit of information. Notice that Bob has received
two photons and the transmission efficiency is one bit per photon.

An esoteric use of entanglement for sharing information between
three parties is presented in \cite{Steane 2000}.

It has been suggested that entanglement may improve the von Neumann
capacity (\ref{eqn 3.2.1}) but this is still an open question.

\section*{Acknowledgment}
Thanks are due to Shlomo Shamai (Shitz) for discussions on the 
concept of capacity per unit cost, 
to Thomas Ericson for helpful comments,
and to Rolf Johannesson who provided the relation (\ref{eqn 5.2}).

\end{document}